\documentclass[prl,twocolumn,preprintnumbers, showpacs,amsmath,amssymb]{revtex4}

\newcommand{\PP}{\textbf P }
\newcommand{\MM}{\textbf M }
\newcommand{\LL}{\textbf L }
\newcommand{\DM}{{DM} }
\newcommand{\DD}{\textbf D }

\usepackage{graphicx}
\usepackage{bbm}
\usepackage{float}

\begin{document}

\preprint{}
\title{Ferroelectrically induced weak-ferromagnetism in a single-phase multiferroic by design}
\author{Craig J. Fennie}
\affiliation{Center for Nanoscale Materials, Argonne National Laboratory,
        Argonne, IL 60439.}

\begin{abstract}
We present a strategy to design structures for which a 
polar lattice distortion induces weak ferromagnetism. We 
identify a large class of multiferroic oxides as potential realizations 
and use density-functional theory to screen several promising 
candidates. By elucidating the interplay between the polarization 
and the Dzyaloshinskii-Moriya vector, we show how the direction 
of the magnetization can be switched between 180$^{\circ}$ 
symmetry equivalent states with an applied electric field.
\end{abstract}

\pacs{75.80.+q,77.80.-e,81.05.Zx }

\maketitle


%
The rational design of new materials with emergent properties is 
a riveting challenge today in materials physics. It begins with 
understanding a mechanism to control the interplay between diverse 
microscopic degrees of freedom in order to create targeted macroscopic 
phenomena and ends with the discovery or design of new material realizations.
When combined with first-principles density-functional theory, this 
approach provides an efficient strategy to survey 
the vast space of possible materials to target for synthesis. 
For example, new multiferroics in which magnetism coexists 
with ferroelectricity have been discovered where magnetic order 
itself induces ferroelectricity. Through this specific spin-lattice interaction, 
it is readily possible to control the direction of the electrical polarization with a
magnetic-field~\cite{max.sang,kimura.review}. 
An equally fundamental but technologically more relevant problem that 
has received far less study is the electric-field control of 
magnetism~\cite{fiebig.nature,ederer.prb.06,zhao.natmat,tokura.prl.07}.
In particular, the electric-field switching of a magnetization between 180$^{\circ}$ symmetry equivalent states
has yet to be demonstrated. The most promising direction 
to achieve this in single-phase materials involves a ferroelectric 
distortion inducing weak-ferromagnetism~\cite{fox.scott,ramesh.spaldin,scott.science.review,scott.nature.review}. 
Discovering a prototypical structure for which this approach might be 
realized, however, has remained elusive.
%

Weak-ferromagnetism (wFM) is the phenomenon whereby the 
predominantly collinear spins of an antiferromagnet cant in such 
a way as to produce a residual magnetization ({\textbf M}). It can arise as a 
relativistic correction to Anderson's superexchange,  
i.e$.$, the Dzyaloshinskii-Moriya (DM) interaction~\cite{dzya,moriya},
$E_{{\rm DM}}$=$D_{ij}$$\cdot$${\textbf S_i}$$\times$${\textbf S_j}$
where  $D$ is the Dzyaloshinskii vector. 
A ferroelectric (FE) distortion can induce wFM when the phenomenological 
invariant $E_{{\rm PLM}}$$\sim${\bf P}$\cdot$({\bf L}$\times${\bf M}) is 
allowed in the energy of the antiferromagnetic-paraelectric (AFM-PE) phase, 
where {\textbf P} and {\textbf L} are the polarization and AFM vector respectively.  
Due to this term a coupling between the sign of \PP and {\textbf M}, 
for  fixed {\textbf L}, is evident. 
It is important to realize that a FE can still exhibit wFM without the 
FE distortion per se causing the wFM, i.e., without $E_{{\rm PLM}}$ 
in the corresponding PE phase. For example, although BiFeO$_3$ 
(the most widely studied multiferroic) can display wFM, such an invariant 
does not exist as previously shown from explicit first-principles 
calculations~\cite{ederer.prb.05} and as we argue below from symmetry. 
The challenge then is understanding how to start with the microscopic properties 
of $E_{{\rm DM}}$ and build a material capturing the macroscopic physics of
$E_{{\rm PLM}}$. In this Letter we present for the first time design criteria 
that facilitate this process.  
In general the criteria target a structure for which wFM is symmetry-forbidden 
in the PE phase but symmetry-allowed in the FE 
phase~\cite{ederer.prb.05,ederer.prb.06}.  
Below we use this to identify a class of multiferroic oxides as 
potential realizations and use density-functional theory to screen 
several promising candidates. One complication arises in magnetic 
systems with broken spatial inversion such as a FE: a symmetry allowed 
interaction~\cite{dzya2} may in some cases generate a long-wavelength 
magnetic spiral canceling the net {\bf M}, e.g., as in (bulk) BiFeO$_3$.  
Here we consider the situation where this inhomogeneous state is 
suppressed, e.g., when single-ion anisotropy is large~\cite{spiral}.

Our strategy is to formulate the problem in terms of a structural-chemical 
criterion and a magnetic criterion. To illustrate the idea we consider a
two-sublattice antiferromagnet such as BiFeO$_3$.  A synopsis of 
the structural-chemical criterion is as follows: 
start with a PE structure, decorate the lattice with spins such that the 
midpoint between two spins coincides with a site having inversion 
($\mathcal{I}$) symmetry (so that $D$ = 0 by symmetry, i.e., Moriya's 
first rule~\cite{moriya}), place a FE-active ion at an $\mathcal{I}$-site. 
Compounds which satisfy this criterion are quite intriguing because if 
there are no other symmetry elements that would forbid wFM -- the magnetic 
criterion -- all that is required to induce a 
non-zero $D$ and wFM is to remove the $\mathcal{I}$ center by 
controlling the off-centering of the FE-active ion  either by temperature, 
pressure, or electric-field. The magnetic criterion is primarily a question 
of how the spins order and of their direction with respect to the 
crystallographic axes.
Notice our microscopic based criteria implies that  \LL  is odd under 
$\mathcal{I}$, which is precisely what is required considering macroscopic 
phenomenology, i.e., $E_{{\rm PLM}}$.
These criteria facilitate evaluating known compounds or when 
combined with crystal chemistry principles designing new prototypes. 
For example, using PE BiFeO$_3$ as a starting point, we apply them
to design a novel structure.

\begin{figure}[t]
 \centering
\includegraphics[width=0.4\textwidth]{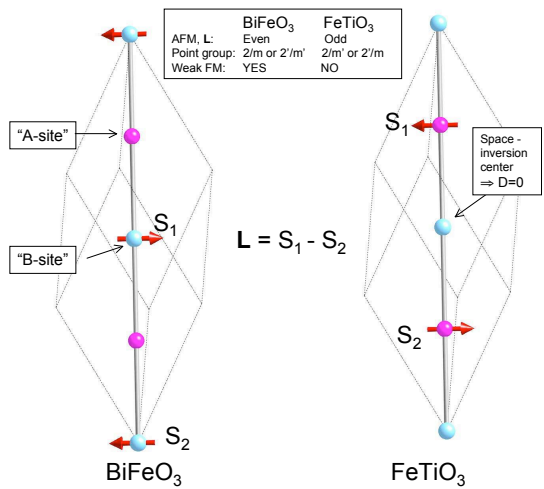}\\
\caption{\label{fig:bifeo3}
Paraelectric ABO$_3$ compounds, space group, R$\bar{3}$c, 
BiFeO$_3$ and FeTiO$_3$. A-site: Wyckoff position 2a, site symmetry 32.
B-site: Wyckoff position 2b site symmetry, $\bar{3}$.}
\end{figure}

In PE BiFeO$_3$, crystallographic space group R$\bar{3}$c,  the Bi-ions 
occupy the $A$-site, Wyckoff position 2a, with local site symmetry 32, 
while the magnetic Fe-ion occupies a site with inversion symmetry, the 
$B$-site, position 2b, site symmetry $\bar{3}$. The Fe-spins order 
ferromagnetically within antiferromagnetically coupled  (111) planes with 
the magnetic easy axis perpendicular to [111]. Although PE BiFeO$_3$ 
satisfies the magnetic criterion, $\mathcal{I}$ transforms each Fe sublattice 
onto itself, $\mathcal{I}${\textbf L} = $\mathcal{I}$($S_1$-$S_2$) = {\textbf L}, 
as can be seen in Fig.~\ref{fig:bifeo3}(a). In this case of $B$-site magnetism 
the structural-chemical criterion is not met and the invariant 
$E_{{\rm PLM}}$ is forbidden by symmetry (the PE point group 
is 2'/m' or 2/m for which wFM is allowed).
In contrast, consider the case where the magnetic ion is on the 
$A$-site and similarly ordered so that the magnetic criterion is still 
satisfied, Fig.~\ref{fig:bifeo3}(b). Now, in this
case of $A$-site magnetism, $\mathcal{I}${\textbf L} = -{\textbf L}. Placing
a FE-active ion such as Ti$^{4+}$ on the $B$-site would then satisfy the structural-chemical
criterion. The PE magnetic point group is now 2/m' (2'/m) in which
wFM is forbidden and {\it by design} a FE distortion, via $E_{{\rm PLM}}$, would lower the symmetry 
to m' (m) thereby inducing wFM.
 The question remaining is whether compounds in our rationalized structure
can be synthesized? As we discuss next, several already exist.


\begin{table}[b]
\caption{Theoretical (experimental)  properties of multiferroic $A$TiO$_3$, space group R3c. $f.u.\equiv$ formula unit.}
\begin{ruledtabular}
\begin{tabular}{lccc}
& Mn (Exp.~\cite{ko})& Fe (Exp.~\cite{ming}) & Ni\\ \hline
a$_h$  & 5.127\AA\,\,(5.205\AA) & 5.05\AA\,\,(5.12\AA)&4.93\AA\\
c$_h$  &13.63\AA\,\,(13.70\AA)  &13.52\AA\,\,(13.76\AA)&13.65\AA\\
$\theta_r$ &56.4$^\circ$\,\,(56.8$^\circ$)  &56.2$^\circ$\,\,(56.0$^\circ$) &54.7$^\circ$\\
T$_{FE}$& $>$1000K & $>$1000K & $>$1000K\\
{\bf P} & 83 $\mu$C/cm$^2$ & 94 $\mu$C/cm$^2$ & 110 $\mu$C/cm$^2$\\
$\Theta_{CW}$ & -290 K & -305 K  & -170 K\\
T$_N$ & 135 K & 260 K & 100 K\\
{\textbf L} & 4.5 $\mu_B$/$f.u.$ & 3.6 $\mu_B$/$f.u.$ & 1.6 $\mu_B$/$f.u.$\\
{\bf M} & -0.002 $\mu_B$/$f.u.$ &  -0.03 $\mu_B$/$f.u.$ & 0.25 $\mu_B$/$f.u.$\\
$J_1$ &-0.5meV &-1.0meV & -1.5meV\\
$K$ &  -0.02meV/$f.u.$ &  -0.74 meV/$f.u.$ &-0.03 meV/$f.u.$ \\
$D$ &-0.0004 meV/$f.u.$ &-0.02 meV/$f.u.$ &-0.35 meV/$f.u.$\\
\end{tabular}
\end{ruledtabular}
\label{table:prop}
\end{table}

The mineral Ilmenite FeTiO$_3$ is one member of a family of 
compounds~\cite{goodenough} that include the titanantes 
$A^{2+}$Ti$^{4+}$O$_3$ with $A$ = Mn--Ni. They are all AFM insulators
with ordering temperatures T$_{\rm N}$$\sim$40K-100K. At atmospheric 
pressure they form in the Ilmenite structure, space group R$\bar{3}$. 
Ilmenite can be thought of as an ordered corundum structure. At 
high-pressure both MnTiO$_3$ and FeTiO$_3$  have been found to 
form a quenchable metastable LiNbO$_3$ LBO-phase, space group 
R3c~\cite{ko,ming}. Note this LBO-phase is structurally isomorphic 
to BiFeO$_3$ except the magnetic and FE atom positions are reversed, 
for example: MnTiO$_3$$\rightarrow$BiFeO$_3$ implies Mn$\rightarrow$Bi 
and Ti$\rightarrow$Fe. This is precisely the structural-chemical criterion
outlined above. The remaining question is to identify the magnetic ground 
state in the FE phase to determine if the magnetic criterion 
is satisfied. In the remainder of this Letter we present a first-principles 
study of the FE and magnetic properties 
of the LBO-phase of MnTiO$_3$, FeTiO$_3$, and NiTiO$_3$. We 
demonstrate that these are realizations of the design criteria and 
provide a novel simple picture of how the interplay of $D$ and \PP leads 
to electric-field control of wFM.

{\it Method.}$-$
We performed density-functional calculations using PAW potentials 
within LSDA+U~\cite{anisimov.jpcm.97} as implemented in {\sf VASP}~\cite{VASP,PAW}. 
The wavefunctions were expanded in plane waves up to a  cutoff of 
500 eV. Integrals over the Brillouin zone were approximated by sums on a 
$6 \times 6 \times 6$ $\Gamma$-centered $k$-point mesh. Phonons were 
calculated using the direct method. Where noted, non-collinear calculations 
with L-S coupling were performed.
To find appropriate values of on-site Coulomb U and exchange J$_{\rm H}$ 
parameters we performed a series of calculations to estimate the Curie-Weiss 
temperature $\Theta_{{\rm CW}}$ as a function of U for MnTiO$_3$, FeTiO$_3$, 
and NiTiO$_3$ in the {\it ground state Ilmenite structure}, space group 
R$\bar{3}$~\cite{heisenberg}. For all compounds a value of U = 4.5 eV 
and J$_{\rm H}$=0.9 eV was found to give a reasonable account of the measured 
values. 
It should be noted that the presented results
do not qualitatively change for reasonable variations of U.

\begin{figure*}[t]
 \centering
\includegraphics[scale=0.6]{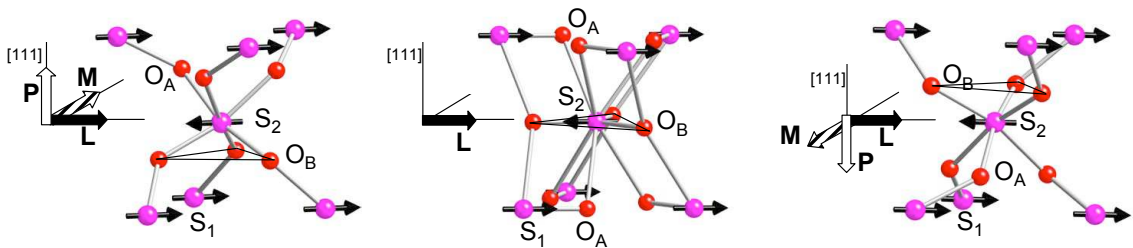}\\
\caption{\label{fig:chiral}
Chiral nature of the $S_1$--O--$S_2$ bonds;
(Left) Polarization-up and (Right) Polarization-down.
(Middle) Paraelectric state showing two equivalent 
 \emph{inter}-planar pathways, 
 e.g$.$ $S_1$--O$_A$--$S_2$  and $S_1$--O$_B$--$S_2$. 
 Note: ${\bf D}\|\PP, \,\,\, \LL \perp \MM \perp {\bf P}.$}
\end{figure*}
%
{\it Ferroelectricity.}$-$ 
In the soft-mode theory of ferroelectricity, the PE to FE transition is 
associated with the softening of a single unstable infrared-active phonon. 
In the R$\bar{3}$c$\rightarrow$R3c transition, this is a PE phonon 
polarized along [111] of symmetry type A$_{2u}$. We calculated the 
frequencies of these A$_{2u}$ phonons at T=0 and found one highly 
unstable mode, e.g$.$, in MnTiO$_3$, $\omega_{soft}$$\approx$i150 
cm$^{-1}$. The character of this soft mode consists of $A$-ion and Ti 
displacements moving against oxygen, similar to other R3c FEs such 
as BiFeO$_3$ and LiNbO$_3$.

Next we broke inversion symmetry and performed a full structural 
optimization within the R3c space group. In Table~\ref{table:prop} the 
calculated lattice constants at T = 0 are shown to be in excellent agreement 
with those observed at room temperature for MnTiO$_3$ and FeTiO$_3$
(R3c NiTiO$_3$ has not yet been synthesized.)
The distortion leading from the PE to the FE structure can be decomposed 
entirely in terms of the soft-mode. The atomic displacements are almost 
equal in magnitude to those in LiNbO$_3$, meeting Abraham's structural 
criteria for switchable ferroelectricity~\cite{abraham}. Based on Abraham's 
empirically derived formula relating the magnitude of these distortions to the 
FE transition temperature~\cite{abraham}, we estimate FE T$_C$$\sim$1500K--2000K.
Finally, using the modern theory of polarization~\cite{king-smith.prb.93} 
we calculated a large \PP $\approx$ 80--100 $\mu$C/cm$^2$  
comparable to that of BiFeO$_3$.

{\it Magnetic structure.}$-$
Weak ferromagnetism arises as a small perturbation - from the relativistic 
spin-orbit interaction - to a predominately collinear magnetic state, i.e$.$,
$|J|$$>>$$|D|$, where $J$ and $D$ are Heisenberg and \DM 
exchange respectively. This difference in energy scales naturally separates 
the problem. As such we first identify the collinear state that minimizes the 
spin-interaction energy without L-S coupling, i.e$.$,
$ E_{\rm H} = -  \sum_{ij} J_{ij} {\bf S}_i \cdot {\bf S}_j, $
by extracting the first four nearest neighbor (n.n$.$) exchange  integrals, $J_n$, 
from total energy calculations~\cite{heisenberg}.
We found the state that minimizes $E_{\rm H}$ consists of spins 
aligned ferromagnetically within antiferromagnetically coupled (111) planes,
consistent with the magnetic criterion outlined above. This magnetic state arises 
due to a strong AFM $J_1$ coupling between n.n$.$ spins in 
adjacent  (111) planes. The N\'eel temperature calculated within mean-field 
theory~\cite{note.neel} was found to be T$_{\rm N}$$\sim$ 100K  for MnTiO$_3$ 
and NiTiO$_3$  and T$_{\rm N}$$\sim$ 250K for FeTiO$_3$.

For a uniaxial crystal the orientation of the global spin axis relative to the 
crystallographic axis is given to lowest order by 
$E_{\rm{ani}}= \sum_i K_i \rm{sin}^2(\theta),$
where $\theta$ is the angle between [111] and {\bf L}. Depending on the 
sign of $K$, the spins lie in a plane perpendicular or parallel to [111], in 
which case wFM is allowed or forbidden, respectively. 
To calculate the single-ion anisotropy constant $K$ we first performed a 
self-consistent density-functional calculation with collinear spins, without 
L-S coupling. Then, using the charge density and wavefunctions, we performed 
a series of non-selfconsistent calculations with spin-orbit interaction included 
for different orientations of the global spin axis.  We found that the magnetic 
easy axis lies in the plane perpendicular to [111] with $K$ ranging from 
-0.03 meV for MnTiO$_3$ and NiTiO$_3$, to -0.7 meV for FeTiO$_3$ (the 
much large anisotropy for FeTiO$_3$ is associated with the orbital degeneracy).
Symmetry then allows an additional contribution to the total energy
which can cause the spins to cant:
$E_{{\rm DM}} =  \sum_{ij} \tilde{D}_{ij} \cdot \left( {\bf S}_i\times {\bf S}_j \right)$,
where $\tilde{D}_{ij}$ is the Dzyaloshinskii vector. Phenomenologically this can 
be described by $E_{{\rm DLM}}$={\bf D}$\cdot$({\bf L}$\times${\bf M}) where 
{\textbf L}=$S_1$-$S_2$ and {\textbf M}=$S_1$+$S_2$. Symmetry requires 
\DD to point along [111], i.e$.$, parallel/antiparallel to {\bf P}. Since $K$$<$0 
requires \LL to be in the plane perpendicular to [111] and subsequently to 
{\bf D}, the induced \MM is perpendicular to {\textbf P}. Once the direction 
of \LL is fixed, the sign of \MM that minimizes $E_{{\rm DLM}}$ is determined 
by the sign of {\bf D}.

Symmetry allows the spins to cant in the FE phase, but do they and by how 
much? To address this question we calculated the self consistent spin-density 
in the presence of the spin-orbit interaction for the FE structure with for example
the polarization pointing down. The spins were initialized in a collinear 
configuration, e.g$.$ {\textbf L}$^0$ = (2g$\mu_B$S, 0, 0), {\textbf M}$^0$ = 
(0,0,0), and then allowed to relax without any symmetry constraints imposed. 
For MnTiO$_3$ the induced moment was rather weak,  \MM= (0, -0.002, 0) 
$\mu_B$/formula unit ($f.u.$), although comparable to the canonical weak-ferromagnet 
Fe$_2$O$_3$ and still measurable. The smallness of this result is not too 
surprising considering that the spin-orbit parameter $\lambda$ vanishes
for Mn$^{2+}$ in the atomic limit. In contrast, $\lambda$ is relatively large 
for Fe$^{2+}$ and Ni$^{2+}$, and correspondingly the induced moments 
increase dramatically; \MM= (0, -0.03,0) $\mu_B$/$f.u.$ and
 \MM=(0,0.25,0) $\mu_B$/$f.u.$ for FeTiO$_3$ and NiTiO$_3$ 
 respectively~\cite{gfactor}.
Finally, we can approximate the strength of $\tilde{D}$ from the calculated 
canting angle and $J_1$.
The results are summarized in Table~\ref{table:prop}.

\begin{figure}[t]
 \centering
\includegraphics[width=0.3\textwidth]{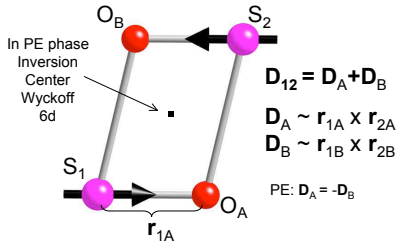}\\
\caption{\label{fig:DM}
Superexchange pathways,  $S_1$-O$_A$-$S_2$ and $S_1$-O$_B$-$S_2$, between 
nearest neighbor spins.
The net Dzyaloshinski-Moriya vector, {\bf D}$_{12}$, contains two contributions,
{\bf D}$_{A}$ and {\bf D}$_{B}$, that cancel in the paraelectric phase.}
\end{figure}

Next we proceed to elucidate the interaction between \MM 
and {\textbf P}.  Similar calculations as we just discussed 
were performed for the PE structure and for the FE structure 
with \PP in the opposite (symmetry equivalent) direction, again 
relaxing the spin-density without symmetry constraints. 
In the former \MM vanished confirming that the FE distortion 
is required for the observed wFM while in the latter \MM switched sign. 
These results are consistent with our earlier symmetry arguments
for a FE inducing wFM, i.e., for  the invariant 
$E_{{\rm PLM}}$$\sim${\bf P}$\cdot$({\bf L}$\times${\bf M})~\cite{plm}
and suggests that the direction of \MM can be switched between 
stable 180$^{\circ}$ directions by an external electric field that would 
switch \PP thereby switching {\bf M}.
One way to understand this result is to argue that the sign of the \DM 
vector \DD depends on the direction of {\bf P}, i.e., \PP $\propto$ {\textbf D}.
At first this may seem puzzling considering the fact that $D_{ij}$ is an 
axial vector. As Ederer and Spaldin pointed out~\cite{ederer.prb.05}, \PP 
would have to change the sense of oxygen rotation 
in order for $D$ to change sign. In Fig.~\ref{fig:chiral} we compare the $S$-O-$S$ 
bonds in the PE and up/down FE states, centering on spin $S_2$ and its nearest
neighbors in adjacent  (111) planes. A change in chirality of the $S$-O-$S$ 
bonds is clearly visible due to a change in 
the direction of {\textbf P}. 
The physics becomes clear by examining the microscopic  $E_{DM}$ and 
realizing that the net Dzyaloshinskii vector
$ \tilde{D}_{ij} = \sum_{\alpha} {D}^{\alpha}_{ij} $
has to be summed over all distinct $S_1$--O$_{\alpha}$--$S_2$ pathways.
In the PE structure two pathways, a left and right chiral, contribute
to $ \tilde{D}_{12}$ as we show in Fig.~\ref{fig:DM}. The orientation of the 
$D^A_{12}$ vector is given by~\cite{keffer,moskvin}
$D^A_{12} \sim r_{1A} \times r_{2A}$ where $r_{1A}$ is the unit vector pointing 
along the $S_1$ - O$_A$ direction. There is, however, an additional pathway 
connecting $S_1$ to $S_2$ through O$_B$. In the PE phase $D^A$ and D$^B$ 
can be shown to have equal magnitude but opposite sign leading to a 
vanishing  net \DM interaction. In the FE phase, \PP strengthens/weakens one 
pathway over the other leading to a finite \DM interaction (in Fig.~\ref{fig:chiral}
we only show the ``strong" $S_1$-O-$S_2$ pathway in the FE phase).
Therefore the origin of ferroelectrically-induced wFM in this class of materials is
a change in the relative contribution of two \DM superexchange pathways (with 
opposite sign) due to a polar lattice distortion.
 %

Today, the challenge in multiferroics has shifted from finding new 
magnetic ferroelectrics to identifying materials in which the polarization 
and the magnetization are strongly coupled. In this work we have presented
criteria that have the potential to advance the 
discovery of such complex materials.
These electrically-controlled switchable magnets provide  fertile 
ground for additional studies of how spin and lattice degrees 
of freedom interact and also hold promise for application
in magnetic devices.

Useful discussions with Matthias Bode, Venkat Gopalan, 
Darrell Schlom, and S.K. Streiffer are acknowledged. Work
at the Center for Nanoscale Materials was supported by US DOE,
Office of Science, Basic Energy Sciences under Contract No. DE-AC02-06CH11357.


\begin{thebibliography}{10}
\bibitem{heisenberg}
Total energy calculations for 10 linearly independent spin
configurations were performed in an 80-atom supercell.
\bibitem{gfactor}
The trend is also consistent with 
  \DD $\sim$ $\Delta$g$J$, where $\Delta$g is the change in 
the g-factor. See Refs. \onlinecite{moriya,goodenough}.
\bibitem{plm}
We performed further calculations with \LL and \PP
initialized in several directions, all results are consistent with 
the symmetry of this interaction, i.e., 
P$_z$(L$_x$M$_y$$-$L$_y$M$_x$).
\end{thebibliography}
\end{document}